\documentclass[conference]{IEEEtran}
\IEEEoverridecommandlockouts

\usepackage{graphicx}
\usepackage{xcolor}
\usepackage{amsmath,bbm}
\usepackage{amssymb}
\usepackage{booktabs}
\usepackage{multirow}

\DeclareMathOperator*{\argmin}{arg\,min}

\begin{document}
\title{Near-Lossless Deep Feature Compression for Collaborative Intelligence \\
\thanks{This work was supported in part by the Vanier Canada Graduate Scholarship and the NSERC Grant RGPIN-2016-04590}}

\author{\IEEEauthorblockN{Hyomin Choi}
\IEEEauthorblockA{School of Engineering Science\\
Simon Fraser University\\
Burnaby, BC, Canada\\
Email: chyomin@sfu.ca}
\and
\IEEEauthorblockN{Ivan V. Baji\'{c}}
\IEEEauthorblockA{School of Engineering Science\\
Simon Fraser University\\
Burnaby, BC, Canada\\
Email: ibajic@ensc.sfu.ca}}

\maketitle
\vspace{-.5cm}

\begin{abstract}
\emph{Collaborative intelligence} is a new paradigm for efficient deployment of deep neural networks across the mobile-cloud infrastructure. By dividing the network between the mobile and the cloud, it is possible to distribute the computational workload such that the overall energy and/or latency of the system is minimized.  %but it is necessary to transfer the feature data obtained from an intermediate layer between segmented networks in order to achieve the network's goal. 
However, this necessitates sending deep feature data from the mobile to the cloud in order to perform inference. 
In this work, we examine the differences between the deep feature data and natural image data, and propose a simple and effective near-lossless deep feature compressor. The proposed method achieves up to 5\% bit rate reduction compared to HEVC-Intra and even more against other popular image codecs. Finally, we suggest an approach for reconstructing the input image from compressed deep features that could serve to supplement the inference performed by the deep model.
\end{abstract}


\section{Introduction}
%With the evolution of deep neural network (DNN), artificial intelligence-enabled mobile and Internet-of-Things (IoT)~\cite{xia2012internet} applications such as intelligent surveillance camera, automated personal assistants, self-driving cars, unmanned aerial vehicles and so on, has been increasing~\cite{poniszewska2018endowing}. By considering battery capacity and cost-effectiveness, or easiness of online, trained DNN engines can be commonly embedded in the cloud server while the terminus devices interact with the server as transmitting required data and receiving the engine's outcome. This approach shifts whole computational complexity on one-side. Conversely, thanks to the advanced graphical processing units (GPUs), mobile/IoT devices could run the engine solely depending on its complexity, then send the output to the sever. However, this approach also loads the full computation on the one side.

Recent advances in deep neural networks (DNNs) are making various artificial intelligence (AI)-enabled 
%and Internet-of-Things (IoT)~\cite{xia2012internet} 
applications feasible: intelligent surveillance cameras, automated personal assistants, self-driving cars, unmanned aerial vehicles, and so on~\cite{poniszewska2018endowing}. A common deployment strategy for AI-based applications on mobile devices is to have the AI model running in the cloud while the terminus device sends its data to it for inference, and receives the results back. In certain cases, small models may run on the terminus device, but the large models that form the backbone of most AI-enabled systems are too power hungry to run on a mobile device. 
%it is By considering battery capacity and cost-effectiveness, or easiness of online, trained DNN engines can be commonly embedded in the cloud server while the terminus devices interact with the server as transmitting required data and receiving the engine's outcome. This approach shifts entire computational complexity to one side. Conversely, it is also possible to carry out the engine on the mobile/IoT devices thanks to the advanced graphical processing units (GPU), but the devices are responsible for the entire computational cost.

A recent study~\cite{kang2017neurosurgeon} proposed a new deployment paradigm called \emph{collaborative intelligence}, whereby a deep model is split between the mobile and the cloud. Extensive experiments under various hardware configurations and wireless connectivity modes revealed that the optimal operating point in terms of energy consumption and/or computational latency involves splitting the model, usually at a point deep in the network. Today's common solutions, where the model sits fully in the cloud or fully at the mobile, were found to be  rarely (if ever) optimal. %contributing to distribution of the computational complexity between the mobile and cloud. Through the diagnosis for the deep neural network in terms of energy consumption and performance latency, \emph{neurosurgeon} operation is performed on the network. %Therefore the network is divided into two parts. %The front layers including the input layer and succeeding layers up to the divided boundary run on the mobile device and is referred to as front-end. 
%Therefore the network is divided into two parts, and the parts are deployed to the mobile and cloud, respectively. Hence, they must collaborate together to conduct the entire network. %remaining layers run on the server. Hence, the mobile and the server collaborate together to achieve the network's goal. 
Another recent study~\cite{jointdnn} extended the notion of collaborative intelligence to model training as well. In this case, data flows both ways: from the cloud to the mobile during back-propagation in training, and from the mobile to the cloud during forward passes in training, as well as inference. 

In these early studies, the issue of compression for the purpose of data transfer between the mobile and the cloud was not studied in detail. In fact,~\cite{kang2017neurosurgeon} assumed the transfer of raw 32-bit floating point feature values, which is rather wasteful. The work in~\cite{jointdnn} included 8-bit quantization followed by PNG coding of quantized feature maps. The work in~\cite{dfc_for_collab_object_detection} was the first to study lossy compression of deep feature data based on HEVC intra coding, in the context of a recent deep model for object detection~\cite{YOLO2}. They noted the degradation of detection performance with increased compression levels and proposed compression-augmented training to minimize this loss by producing a model that is more robust to quantization noise in feature values. However, this is still a sub-optimal solution, because the codec employed is highly complex~\cite{hevc_complexity_analysis} and optimized for natural scene compression rather than deep feature compression.

%inference and forward as well both intensified the collaboration by allowing data flow both ways to intervene training phase and optimize inference accuracy. Furthermore, the study emphasized the necessity for compressing data to transfer because it turns out that the communication process between the mobile and cloud consumes over 75\% of the entire energy and delay cost. The amount of the cost is highly correlated with the feature data volume. Therefore,~\cite{jointdnn,dfc_for_collab_object_detection} examined the compression efficiency for the extracted feature data from the front-end boundary layer. However, these studies na{\"i}vely employed conventional compression algorithms which are thoroughly tailored to natural image/video data and considerably complex~\cite{hevc_complexity_analysis}. With the complicated compression may adversely increase the energy and delay cost rather than reducing them.

A related work~\cite{deepsic} presented semantic image compression by encoding deep features and then reconstructing the input image from them. The compression was based on uniform quantization followed by context-based adaptive arithmetic coding (CABAC) from H.264~\cite{CABAC_H264}. This work was positioned as an image codec that preserves semantic information for image classification, rather than a tool for collaborative intelligence, but the similarities are evident. Although the overall compression efficiency of this approach was somewhat lower than JPEG and JPEG2000, the authors argued that the benefits lie in better preservation of semantic information.  

%Although the main priority in the collaborative approach is to save transmission cost by compressing the intermediate feature data, but in case of coding multimedia data such as image/video and audio signals, inferring the input sources from the transferred data at the cloud server can be also useful in terms of quality of service (QoS) point of view. A recent study~\cite{deepsic} released a jointly optimized network for both compression (including compressed bitrate and reconstructed input distortion) and discrimination information in the loss function. However, this joint-optimization approach has difficulties to apply to pre-trained and -deployed networks. Moreover, it showed that their designed network noticeably loses discrimination accuracy which is the top priority of the network by increasing compression ratio, although the reconstructed input holds visually good quality which is minor priority. (It has not been directly studied that how the joint-optimization affects existing networks when it combines together). To our knowledge, this is the only literature study coping with both feature compression and inferring the input data together in the \emph{collaborative intelligence} approach.

With a view towards collaborative intelligence, in this work we propose a simple and effective near-lossless compression method tailored to deep feature data. We focus on deep models for object detection~\cite{YOLO2} and image classification~\cite{vgg16}, but the approach is applicable to other deep models as well. %based on the deep feature data analysis using well-known discrimination networks in the collaborative approach.
In Section~\ref{sec:data_analysis} we analyze feature data from the two models under study, and note some of the statistical differences between deep feature data and input image data. This analysis informs the design of the proposed compression scheme in Section~\ref{sec:compression}. Furthermore, we demonstrate the capability of reconstructing the input image from compressed deep features in Section~\ref{sec:reconstruction} by constructing and training a mirror model of front-end layers. Experimental results and conclusions are presented in Sections~\ref{sec:experiments} and~\ref{sec:conclusion}, respectively.

\section{Deep feature data analysis}
\label{sec:data_analysis}

\begin{figure}[t]
    \begin{minipage}[b]{1.0\linewidth}
    \centering
    \includegraphics[width=\textwidth]{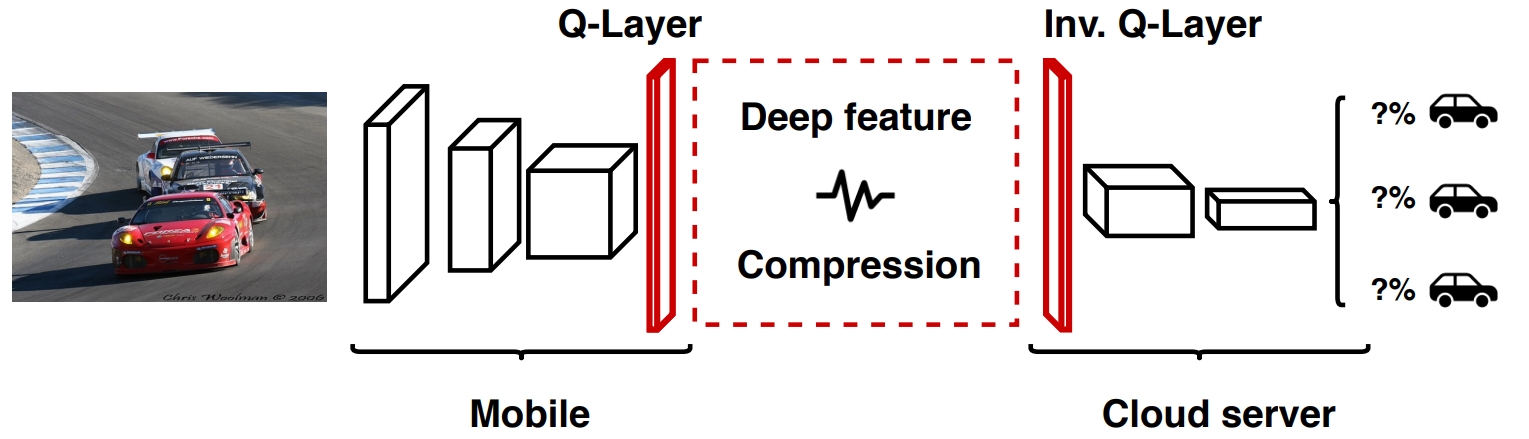}
    \end{minipage}
\vspace{-0.6cm}
\caption{Collaborative intelligence approach with deep feature compression}
\label{fig:dfc_collab}
\vspace{-.3cm}
\end{figure}

\begin{figure*}[t]
    \begin{minipage}[b]{0.32\linewidth}
    \centering
    \includegraphics[width=\textwidth, height=0.5\textwidth]{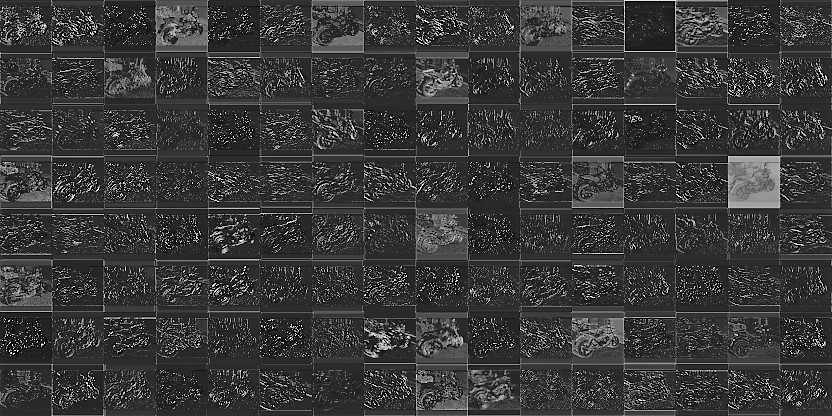}
    \centerline{(a)}\medskip
    \end{minipage}
    \hspace{0.05cm}
    \begin{minipage}[b]{0.32\linewidth}
    \centering
    \includegraphics[width=0.5\textwidth, height=0.5\textwidth]{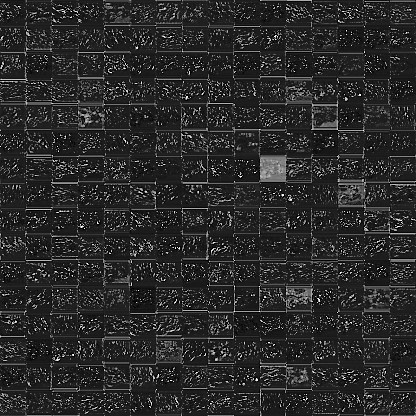}
    \centerline{(b)}\medskip
    \end{minipage}
    \hspace{0.05cm}
    \begin{minipage}[b]{0.32\linewidth}
    \centering
    \includegraphics[width=\textwidth, height=0.5\textwidth]{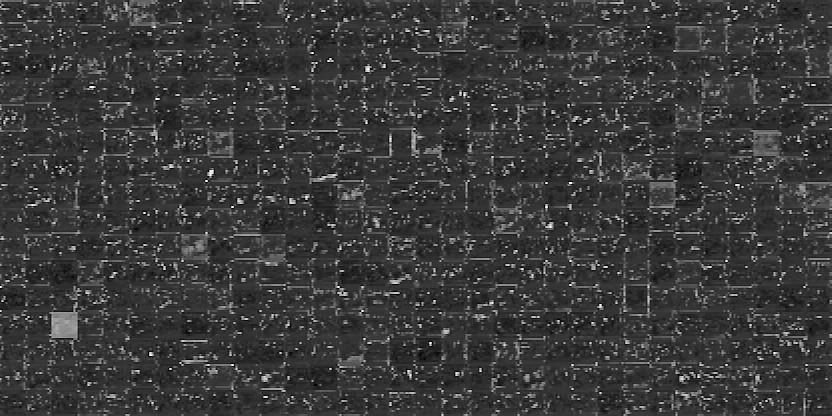}
    \centerline{(c)}\medskip
    \end{minipage}

    \begin{minipage}[b]{0.32\linewidth}
    \centering
    \includegraphics[width=\textwidth, height=0.5\textwidth]{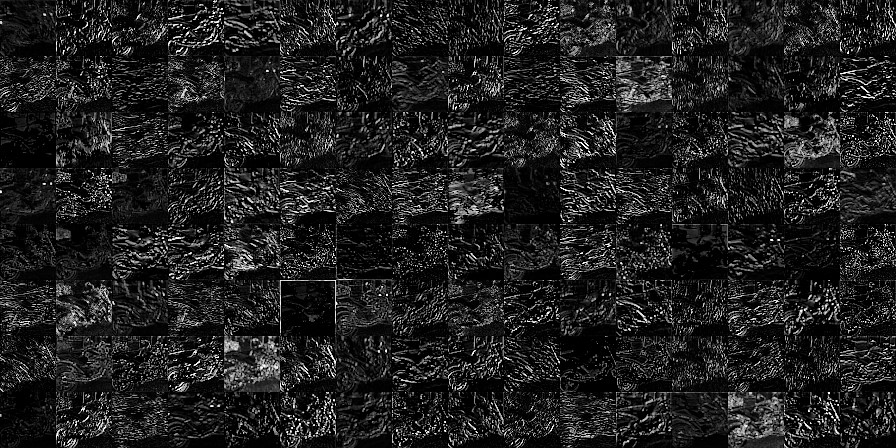}
    \centerline{(d)}\medskip
    \end{minipage}
    \hspace{0.05cm}
    \begin{minipage}[b]{0.32\linewidth}
    \centering
    \includegraphics[width=0.5\textwidth, height=0.5\textwidth]{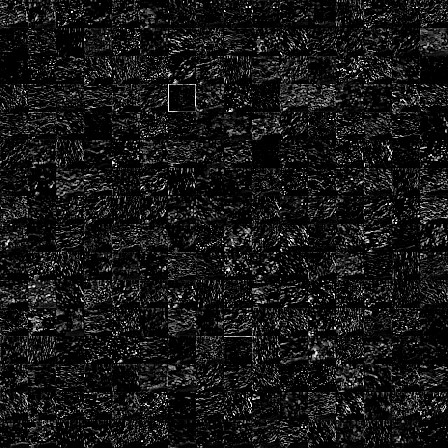}
    \centerline{(e)}\medskip
    \end{minipage}
    \hspace{0.05cm}
    \begin{minipage}[b]{0.32\linewidth}
    \centering
    \includegraphics[width=\textwidth, height=0.5\textwidth]{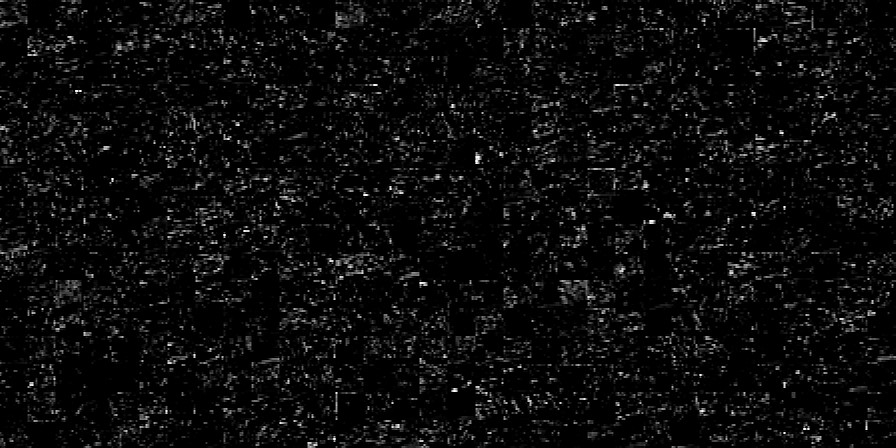}
    \centerline{(f)}\medskip
    \end{minipage}
\vspace{-.2cm}
\caption{Quantized deep features (enhanced for visualization purposes) from YOLOv2~\cite{YOLO2} and VGG16~\cite{vgg16} at various points in the network. Top row: (a) seventh %(832$\times$416), 
(b) eleventh %(416$\times$416) 
and (c) seventeenth %(416$\times$208) 
layer in YOLOv2. Bottom row: (d) sixth %(896$\times$448),
(e) tenth %(448$\times$448) 
and (f) fourteenth %(448$\times$224) 
layer in VGG16. %Note that all output have been enhanced for visualization purposes. Especially, (c) and (f) are also properly scaled.
}
\label{fig:deep_features}
\vspace{-.1cm}
\end{figure*}

\begin{figure*}[t]
    \begin{minipage}[b]{0.32\linewidth}
    \centering
    \includegraphics[width=\textwidth]{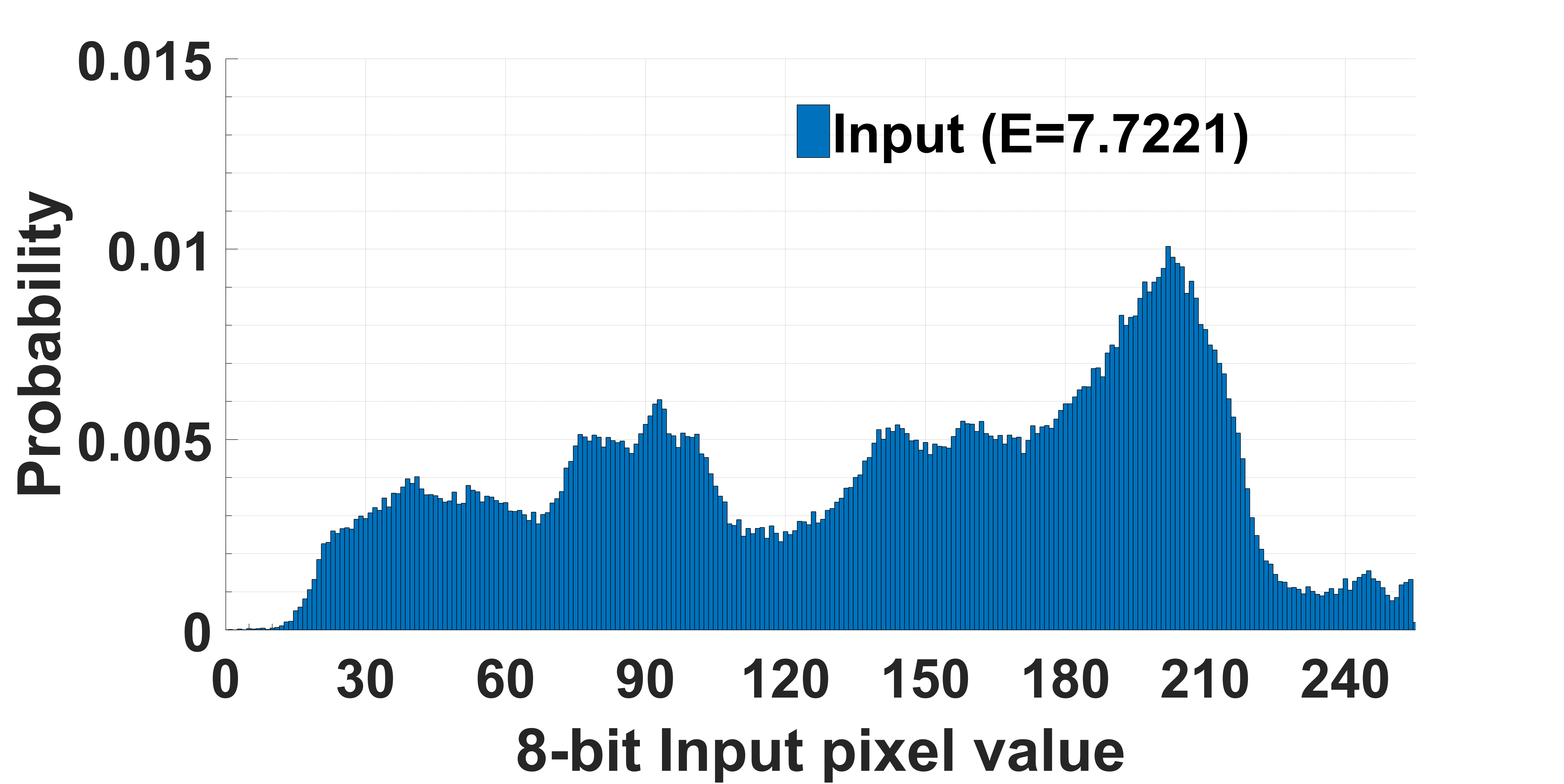}
    \centerline{(a)}\medskip
    \end{minipage}
    \hspace{0.05cm}
    \begin{minipage}[b]{0.32\linewidth}
    \centering
    \includegraphics[width=\textwidth]{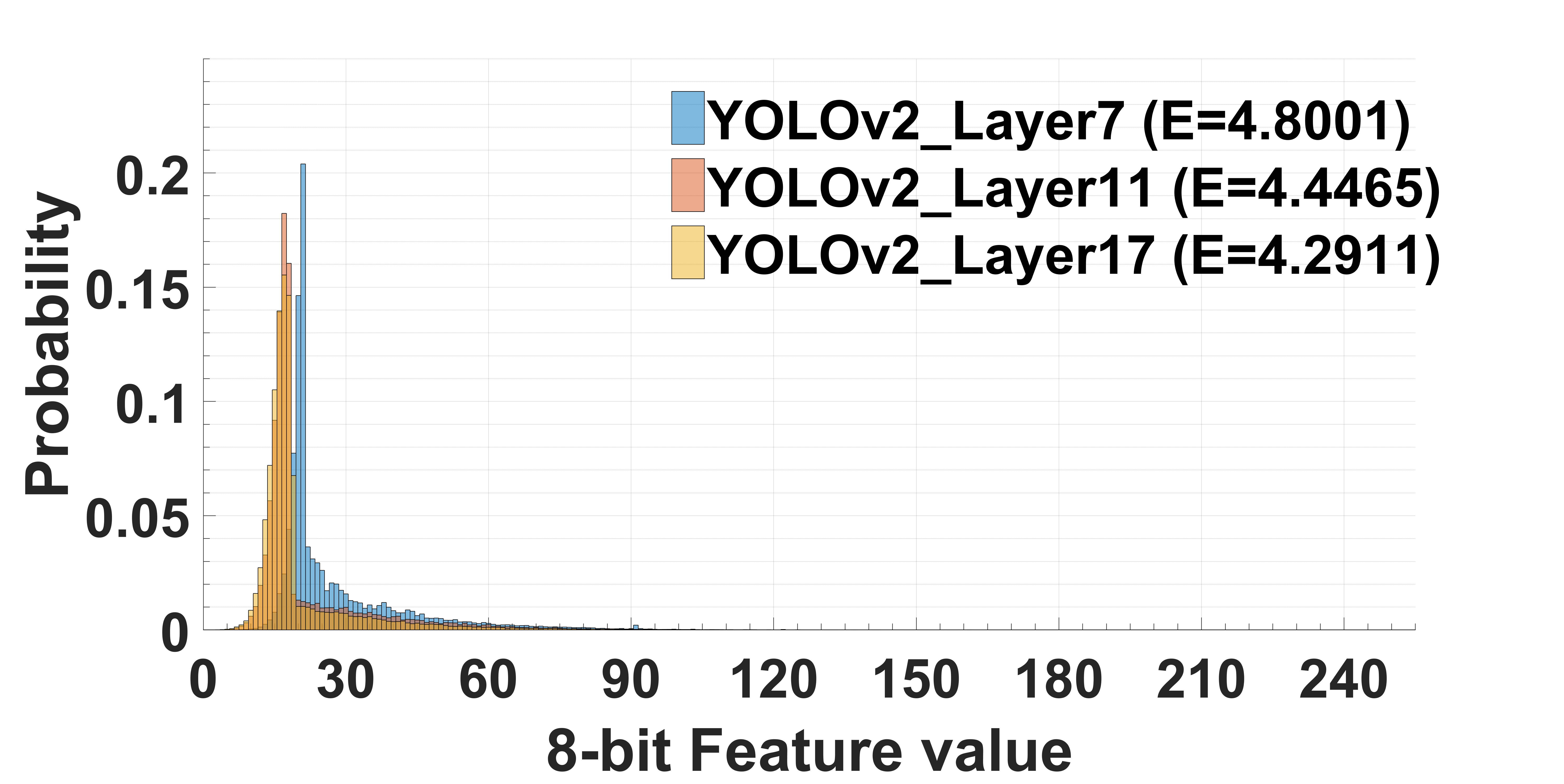}
    \centerline{(b)}\medskip
    \end{minipage}
    \hspace{0.05cm}
    \begin{minipage}[b]{0.32\linewidth}
    \centering
    \includegraphics[width=\textwidth]{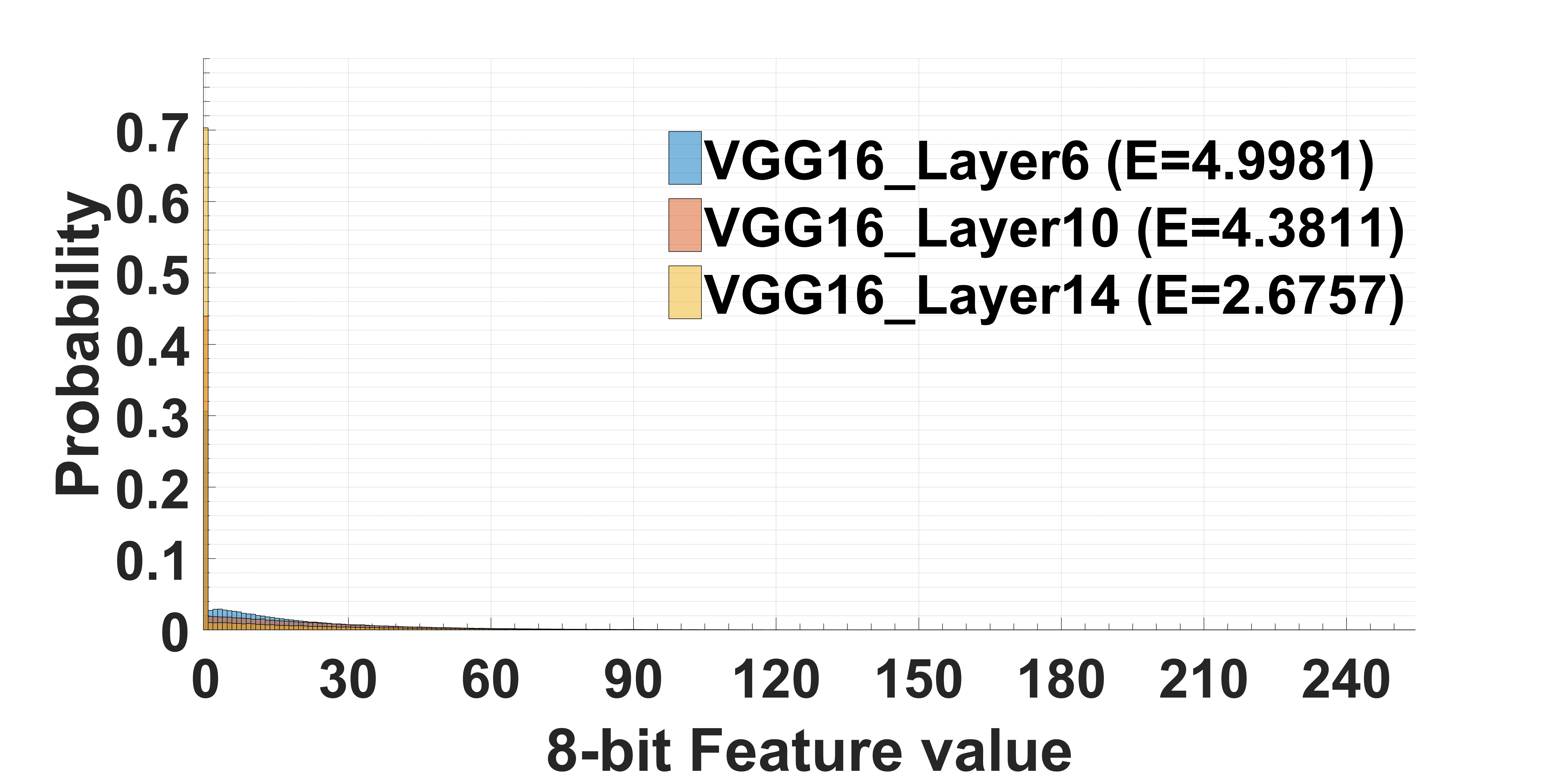}
    \centerline{(c)}\medskip
    \end{minipage}

    \begin{minipage}[b]{0.32\linewidth}
    \centering
    \includegraphics[width=\textwidth]{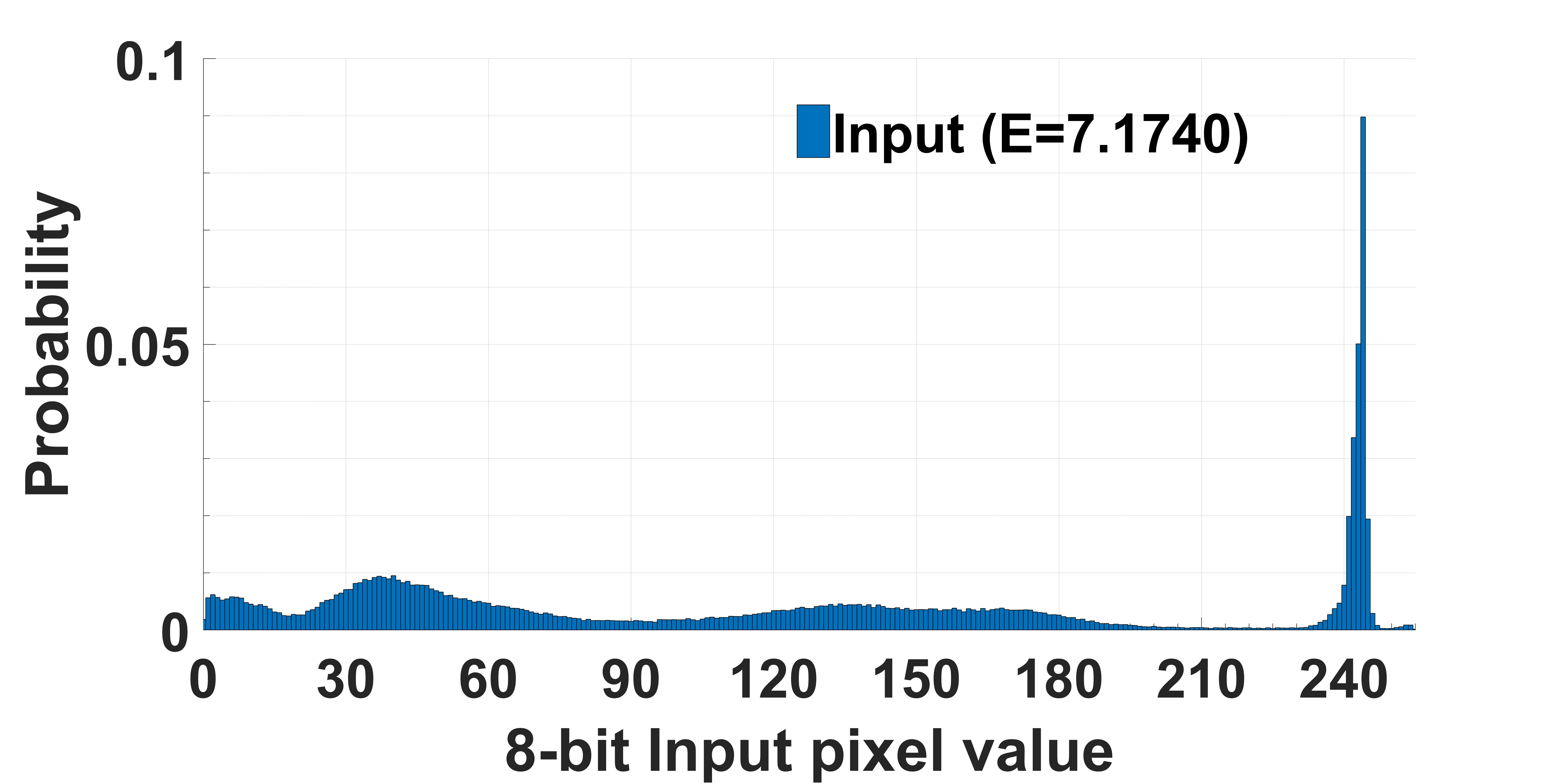}
    \centerline{(d)}\medskip
    \end{minipage}
    \hspace{0.05cm}
    \begin{minipage}[b]{0.32\linewidth}
    \centering
    \includegraphics[width=\textwidth]{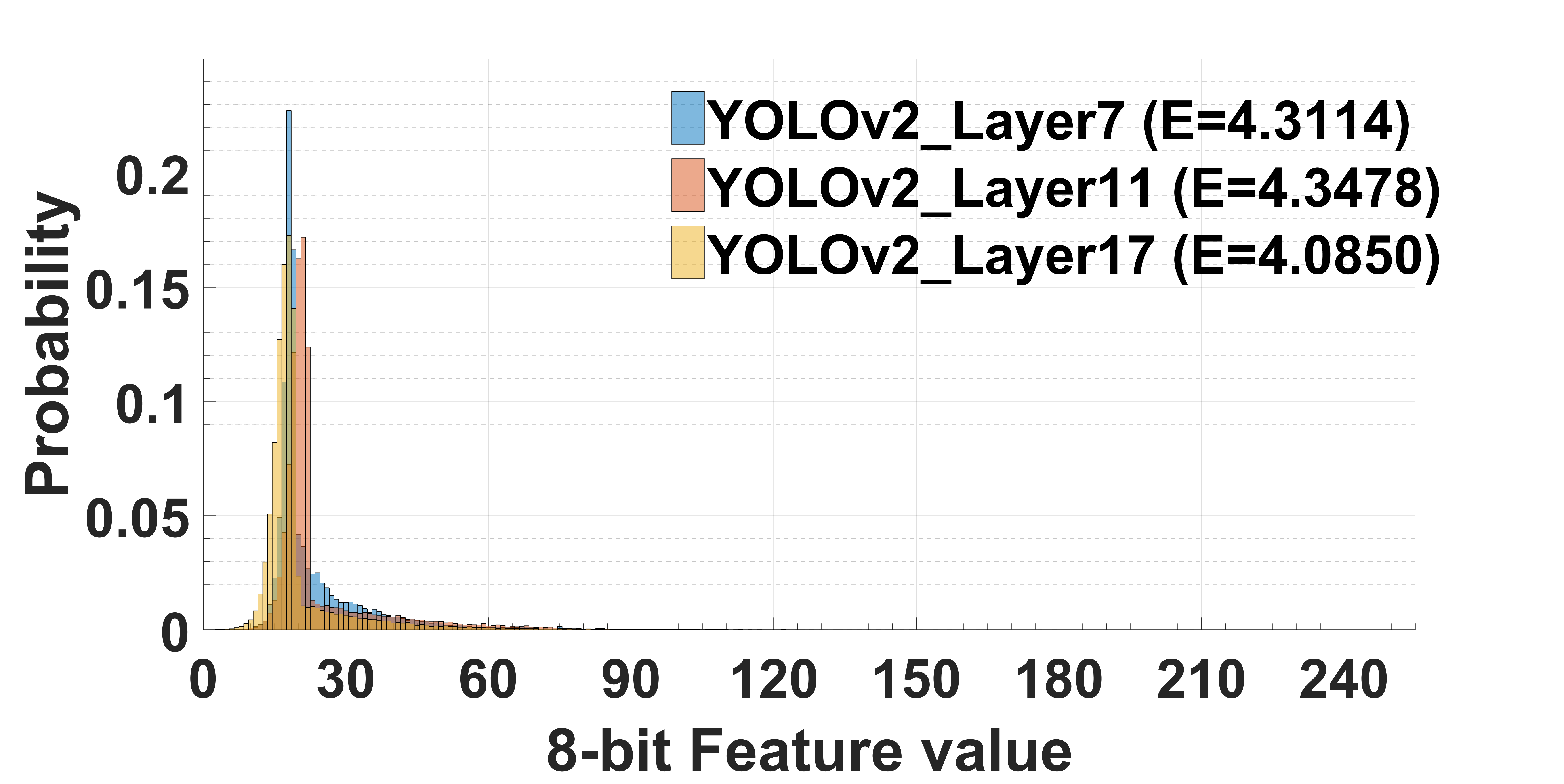}
    \centerline{(e)}\medskip
    \end{minipage}
    \hspace{0.05cm}
    \begin{minipage}[b]{0.32\linewidth}
    \centering
    \includegraphics[width=\textwidth]{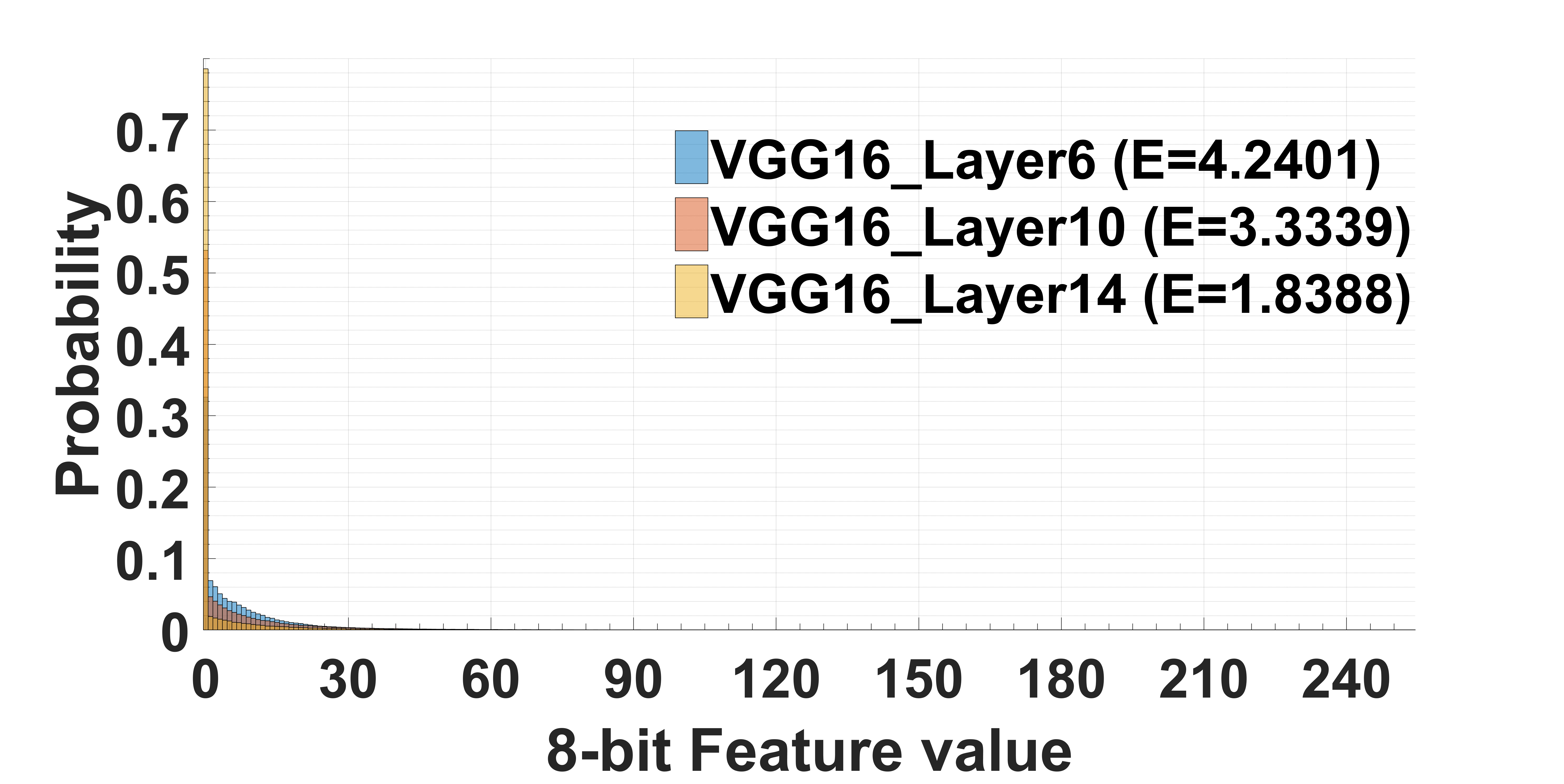}
    \centerline{(f)}\medskip
    \end{minipage}
\vspace{-.2cm}
\caption{(a) and (d) are histograms of pixel values of two input images, (b) and (e) are histograms of the quantized output of the seventh, eleventh and seventeenth layer of YOLOv2, while (c) and (f) are histograms of the quantized output of the sixth, tenth and fourteenth layer of VGG16. Corresponding entropies (E, in bits) are indicated in the plots.}
\label{fig:deep_feature_histogram}
\vspace{-.2cm}
\end{figure*}

Fig.~\ref{fig:dfc_collab} shows the basic setup for collaborative intelligence, where the feature data produced by the initial layers of the deep model are compressed and sent to the cloud for further processing. The efficiency of this approach lies in the fact that for many deep models based on convolutional neural networks (CNNs), the feature data volume (i.e., the total number of feature values) decreases as we move deeper into the network~\cite{kang2017neurosurgeon,jointdnn,dfc_for_collab_object_detection}. Feature values are typically quantized using an $n$-bit uniform quantizer (Q-layer in Fig.~\ref{fig:dfc_collab}) prior to lossless~\cite{jointdnn} or lossy~\cite{dfc_for_collab_object_detection} compression.  %because the feature data decreases as the network data moves toward the end of the network.  Quantizing the layer's output with a certain bit range is more efficient than using 32-bit floating points in terms of transferring and storing the feature data unless there is a significant network's performance drop. %Considering a general network training using wide dynamic range, quantizing the output with a certain bits range is more efficient for either transferring or storing purposes unless there is a significant network's performance drop.
\begin{equation} 
\widetilde{\mathbf{V}} = \textup{round} \left (\frac{\mathbf{V}-\min(\mathbf{V})}{\max(\mathbf{V})-\min(\mathbf{V})}\cdot (2^{n}-1)\right )
\label{eq:quantized_vector}
\end{equation}
\noindent where $\mathbf{V} \in \mathbb{R}^{N \times M \times C}$ is the feature tensor with  $N$ rows, $M$ columns, and $C$ channels at the point of split, $\widetilde{\mathbf{V}}$ is the quantized feature tensor, and $\min(\mathbf{V})$ and $\max(\mathbf{V})$ are the minimum and maximum value in $\mathbf{V}$, respectively. 
%Often, $C$ is a power of $2$ for exploiting parallelism on a GPU. 
In the studies performed so far~\cite{jointdnn,dfc_for_collab_object_detection,deepsic}, this uniform $n$-bit quantization was shown to have negligible effect on image classification and object detection accuracy, for $n\geq6$. For this reason, when such uniform quantizer is followed up by a lossless encoder, we refer to the resulting approach as \emph{near-lossless compression}. In this work, the Q-layer performs uniform $8$-bit quantization. 
%Despite the fact that the quntization process can lose some of the feature data, previous studies have shown that accuracy loss by quantizing the data with 6-, 8- or more bit is negligible. Hence, we also apply the quantization (Q) and inverse Q layers with 8-bit precision to the end of the front-end and the beginning of the back-end, respectively. We present the data as 
%$n_{bit}$ is equal to 8 and round$(\cdot)$ performs rounding to the nearest integer. 
Note that $\min(\mathbf{V})$ and $\max(\mathbf{V})$ need to be transferred to the cloud for the inverse Q process. 

The quantized features $\widetilde{\mathbf{V}}$ are rearranged in a tiled image, as shown in Fig.~\ref{fig:deep_features}. With $C$ channels, we place  %$1 << (\textup{ceil}(\log_{2}C >> 1))$ feature channel width-wise, and $1 << (\textup{floor}(\log_{2}C >> 1))$ height-wise. 
\begin{equation} 
2^{\textup{ceil}\left(\frac{1}{2}\log_{2}C\right)} \qquad \text{and} \qquad 2^{\textup{floor}\left(\frac{1}{2}\log_{2}C\right)}
\label{eq:tiling}
\end{equation}
feature channels (tiles) width-wise and height-wise, respectively. Here, ceil$(\cdot)$ and floor$(\cdot)$ represent ceiling and flooring to the nearest integer, respectively.

Fig.~\ref{fig:deep_features} shows the tiled quantized features obtained from  YOLOv2~\cite{YOLO2} and VGG16~\cite{vgg16}  with default weights for the same input image. %We consider these to be representatives of deep models for object detection and image classification, respectively. 
For each model, three layers are selected (max-pooling layers in all cases) where the resulting feature volume is of comparable size between the two models. We see that, qualitatively, the features are different between the two models, which is not surprising considering that they were trained for different purposes. There are several tiles that still contain somewhat interpretable structures in  Fig.~\ref{fig:deep_features}(a) and (d), but the features become more abstract as we move towards deeper layers (b), (c), (e) and (f). Both sets of features also differ significantly from natural images.   
%~\cite{YOLO2} is the latest single-pass object detection network and~\cite{vgg16} is the representative image classification network. Specifically, 
%These networks have different design and configuration for training such as number of filters and size for the convolutional layers, activation function type, learning rate, data augmentation strategy and etc. All of these factors contribute to training the network parameters with different values, thus the each network produces completely different deep features as shown in fig.~\ref{fig:deep_features}. %The features are tiled within a frame regarding each layer's $N$, $M$ and $C$. 
%We choose examination layers by comparing feature data volumes with average input data size (Too shallow or deep layers are avoided). Also the layers from each network are selected from comparably similar depth. Therefore, we obtained the features from the output of the selected layers, which are all max-pooling layer. Note that acquiring the features from the output of the max-pooling layer always reduces the feature data volume, which is advantageous point. 

Fig.~\ref{fig:deep_feature_histogram} shows pixel and feature histograms for two different input images. %compares histograms of the two input sources' pixels with three layers' outputs of each network as examples. 
As shown in Fig.~\ref{fig:deep_feature_histogram}(a) and (d), which present the luma histograms of input images, pixel intensities in natural images tend to be distributed over the entire range 0-255. In these two input images, no pixel value has a probability above $0.1$. %Meanwhile, the pixel probabilities of input sources are widely distributed over full range of 8-bit precision. There is a dominantly soared point around the certain pixel value of 240 in (d), but the probability of the point is less than 0.1. 
Meanwhile, histograms of quantized feature values from all channels in Fig.~\ref{fig:deep_feature_histogram}(b), (c), (e), and (f) are much more concentrated, and they tend to become more concentrated as we move deeper into the network. Entropy values indicated in figure legends confirm this quantitatively. 

While feature value concentration occurs in both YOLOv2 and VGG16, it is interesting that the concentration points in these two models are different. VGG16 uses Rectified Linear Unit (ReLU) activations~\cite{Goodfellow-et-al-2016}, which are lower-bounded by zero, and the resulting features values concentrate near zero. On the other hand, YOLOv2 uses leaky ReLU activations~\cite{Goodfellow-et-al-2016}, which admit negative values. Hence, prior to quantization, feature values concentrate near zero, but negative feature values exist (i.e., $\min(\mathbf{V})<0$). After quantization~(\ref{eq:quantized_vector}), zero gets mapped to a small positive value (usually 15-25), so the concentration point of quantized features is away from zero. 

%On the other hand, the output features of the intermediate layers are concentrated in a specific range in spite of the 8-bit precision [0-255]. Fig.~\ref{fig:deep_feature_histogram}(b) and (e) shows that the output features from~\cite{YOLO2} are localized to the non-zero values and its locations vary depending on the layer and input sources. For the histogram of the output features from~\cite{vgg16} in fig.~\ref{fig:deep_feature_histogram}(c) and (f), most output features are extremely concentrated on zero. Unlike the natural image data, the deep feature data shows the significantly limited data range. The filtering process of the convolutional layer would be the reason by highlighting the high-correlation points but suppressing otherwise. Moreover, it tends to intensify the effect at the deeper layers, especially at the classification problem. 

\begin{table}[b]
\centering
\vspace{-0.4cm}
\caption{Similarity of pixel/feature values with spatial neighbors and most frequent values}
\vspace{-0.2cm}
\label{tbl:neighbour similarity}
\smallskip\noindent
\resizebox{\linewidth}{!}{%
\begin{tabular}{@{}lcccccccc@{}}
\toprule
                          &      & $\textup{AD}_{m}^2$ & $ \textup{AD}_{l}^2 $ & $\textup{AD}_{t}^2 $ & $\textup{AD}_{tl}^2 $ & $ \textup{AD}_{bl}^2  $ & $\textup{AD}_{tr}^2 $ & none             \\ \midrule
\multicolumn{2}{l}{Input image} & \textbf{22.06\%}                                           & \textbf{26.19\%}                    & \textbf{10.77\%}                    & 4.83\%                               & 4.23\%                               & 3.27\%                               & \textbf{28.65\%} \\ \midrule
\multirow{3}{*}{YOLOv2}  & L7   & \textbf{67.59\%}                                           & 9.75\%                              & 3.65\%                              & 1.27\%                               & 1.20\%                               & 1.02\%                               & \textbf{15.52\%} \\ \cmidrule(l){2-9} 
                          & L11  & \textbf{73.81\%}                                           & 5.80\%                              & 2.34\%                              & 0.81\%                               & 0.80\%                               & 0.71\%                               & \textbf{15.73\%} \\ \cmidrule(l){2-9} 
                          & L17  & \textbf{83.63\%}                                           & 1.12\%                              & 0.30\%                              & 0.28\%                               & 0.28\%                               & 0.25\%                               & \textbf{14.14\%} \\ \midrule
\multirow{3}{*}{VGG16}  & L6   & \textbf{64.49\%}                                           & 6.69\%                              & 4.30\%                              & 1.86\%                               & 1.82\%                               & 1.56\%                               & \textbf{19.28\%} \\ \cmidrule(l){2-9} 
                          & L10  & \textbf{69.62\%}                                           & 3.75\%                              & 2.72\%                              & 1.42\%                               & 1.41\%                               & 1.26\%                               & \textbf{19.82\%} \\ \cmidrule(l){2-9} 
                          & L14  & \textbf{83.68\%}                                           & 1.12\%                              & 0.84\%                              & 0.45\%                               & 0.46\%                               & 0.43\%                               & \textbf{13.02\%} \\ \bottomrule
\end{tabular}}
\end{table}

Next we examine spatial statistics. Specifically, we look at the similarity between the current pixel and its neighbors: left ($l$), top ($t$), top-left ($tl$), bottom-left ($bl$), and top-right ($tr$). We also look at the similarity with the 8 most frequent values in a given histogram: $m_{i}, i=0,1,...7$. To capture the similarity, we consider indicators $\textup{AD}_k^T$ for a given threshold $T$, where $k \in \left\{m, l, t, tl, bl, tr \right\}$. $\textup{AD}_m^T$ is incremented if the absolute difference between the current pixel/feature value $x$ and any (one or more) of the $m_i$ values is less than $T$: $|x-m_i|<T$ for any $i$. If $|x-m_i|\geq T$ for all $i$, then we test the similarity with $k \in \left\{l, t, tl, bl, tr \right\}$, in that order. If $|x-k|<T$, $\textup{AD}_k^T$ is incremented and we move to the next pixel/feature value. Table~\ref{tbl:neighbour similarity} shows $\textup{AD}_k^T$ for $T=2$, expressed as percentages. %how similar the current pel, $P_{c}$, is to the neighbours, left($l$), top($t$), top-left($tl$), bottom-left($bl$), top-right($tr$), and eight most frequent feature pels($m_{i}$) where $i=0,1,...7$. $\textup{AD}_{k}^{2}$ presents that absolute difference between $P_{c}$ and the neighbour $P_{k}$ is less than or equal to 2, $\textup{AD}_{k}^{2} = \left| P_{c} - P_{k} \right| \leq 2$, where $k \in \left\{ m, l, t, tl, bl, tr \right\}$. We examined the similarity check process in the order as described in the first row of the table. So, if the current pel satisfies the similarity condition with the first candidate, then it would not be counted in the subsequent conditions. For the eight most frequent pels, it is counted if the difference with any one of the eight pels is less then or equal to 2. %Note that there is no duplicated counts and it is examined in the priority order.
The results were obtained on the 2510 images from the VOC2007 dataset~\cite{pascal-voc-2007}. Compared to the natural image statistics (second row in the table), we note that feature values exhibit much more similarity with the most frequent values ($\textup{AD}_m^2$), and much less similarity with spatial neighbors ($\textup{AD}_k^2$, for $k \in \left\{l, t, tl, bl, tr \right\}$). This trend increases as we move deeper into the network. Hence, one cannot expect that natural image codecs, which place strong emphasis on spatial redundancy, would be optimal for encoding deep feature data -- new approaches are needed for this purpose.  
%which are input sources, the deep feature data from the middle of networks show the high similarity with the eight most frequent pels mostly rather than any single neighboring pel. In contrast, especially left and top neighboring pixels show the high correlation with the current pixel in the natural image.

\section{Deep feature compression}
\label{sec:compression}
\begin{figure}[t]
    \begin{minipage}[b]{1.0\linewidth}
    \centering
    \includegraphics[width=\textwidth]{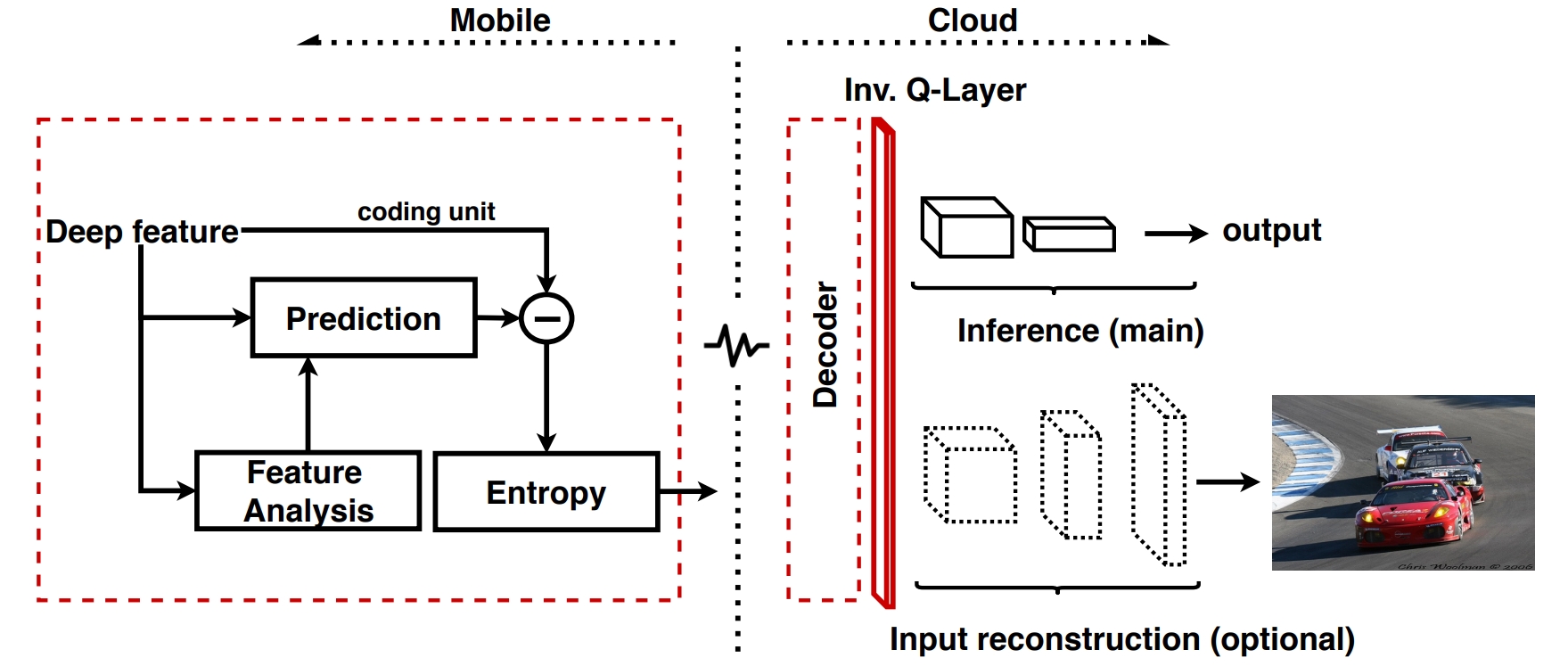}
    \end{minipage}
%\vspace{-0.6cm}
\caption{Proposed deep feature compression for collaborative intelligence}
\label{fig:proposal}
\vspace{-.3cm}
\end{figure}

Fig.~\ref{fig:proposal} shows the proposed compression framework for deep features in  collaborative intelligence applications. In the cloud, deep features are decoded and used for inference. They can also optionally be used for input reconstruction, as discussed in Section~\ref{sec:reconstruction}.   
Before coding the quantized feature data, the following parameters are encoded directly using fixed-length coding: dimensions of the feature tensor, $\min(\mathbf{V})$ and $\max(\mathbf{V})$ (32-bit each) and the eight most frequent feature values, $m_{i}$ for $i=0,1,...7$. The set of $\{m_i\}$ is obtained over the entire quantized feature tensor. A vector of these values, $\mathbf{p}=(p_0, p_1, ..., p_7)$, is referred to as the \emph{palette} vector. 

Initially, the palette vector is  sorted according to the frequency of these values in the first tile,   %\textcolor{red}{Since the initial palette state, $(p_0, p_1, ..., p_7)$, in the header is sorted in advance according to their frequency in the first tile, }
%Initially, \textcolor{red}{as the sorted} $\{m_i\}$ \textcolor{red}{in the header according to their frequency in the first tile are signalled,} %are sorted according to their frequency in the first tile, 
so that 
$p_0$ is the most frequent of the $m_i$'s in the first tile, $p_1$ is the next most frequent, etc. %This is the initial palette state. %These values are then encoded according to Table~\ref{tbl:mfp_table}, which represents the truncated unary code with 8 symbols~\cite{sze2014entropy}.
As we move to other tiles, the palette vector $\mathbf{p}=(p_0, p_1, ..., p_7)$ is re-sorted according to the frequency of occurrence of $m_i$'s up to the previously coded tile,  
so that $p_0$ is the most frequent $m_i$ up to that point, and so on. At the tile boundary, once $\mathbf{p}$ is updated, one element of $\mathbf{p}$ is chosen to minimize the mean absolute difference (MAD) from the feature values in the to-be-coded tile. Its index is found as
\begin{equation} 
j^{*} = \argmin_{0 \leq j \leq 7}\sum_i{|x_i-p_j|}
\label{eq:palette_value}
\end{equation}
where $i$ goes over all locations in the tile. Once $j^{*}$ is found, it is encoded using the truncated 8-symbol unary code~\cite{sze2014entropy} shown in Table~\ref{tbl:mfp_table}. %Palette state and the corresponding codewords in Table~\ref{tbl:mfp_table} are then used to indicate some of the prediction decisions, as described next.   

%However, pels of the input deep feature, but the coded order is decided by only counting the $P_{m_{i}}$'s frequencies within the first tile feature. Note that the size of a tile is multiple of four with reference to unit coding size of 4$\times$4. If the channel size is a multiple of 4, the tile size is same as the channel size. Afterwards, the order is updated by re-computing the frequencies over the previously coded regions. For each tile, one $P_{m}^{n}$, where $n$ is the tile number, is used for the prediction process. We choose the $P_{m}^{n}$ which has the minimum absolute different with the feature pels in the tile. Then, the symbol's order is coded by using variable length coding shown in the table~\ref{tbl:mfp_table}. The binary(bin) are derived by the truncated unary algorithm with maximum 8 symbols~\cite{sze2014entropy}.

\begin{table}[t]
\centering
\caption{Unary code for the palette index}
\label{tbl:mfp_table}
\begin{tabular}{@{}cccc@{}}
\toprule
Index $j^{*}$ & Codeword & Index $j^{*}$ & Codeword  \\ \midrule
$0$        & 1      & $4$        & 11110   \\
$1$        & 10     & $5$        & 111110  \\
$2$        & 110    & $6$        & 1111110 \\
$3$        & 1110   & $7$        & 1111111 \\ \bottomrule
\end{tabular}
\vspace{-.4cm}
\end{table}

\begin{figure}[t]
    \begin{minipage}[b]{1\linewidth}
    \centering
    \includegraphics[width=0.3\textwidth]{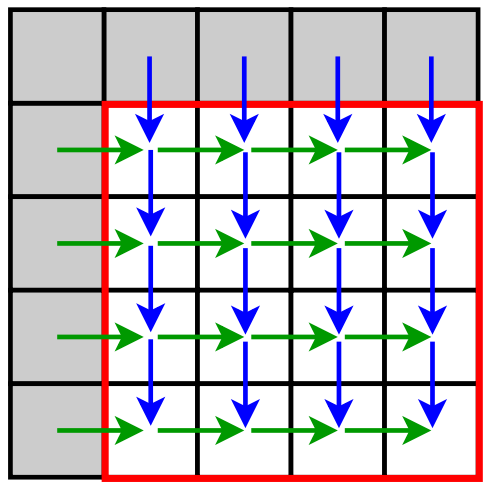}
    \end{minipage}
%\vspace{-0.6cm}
\caption{Illustration of \texttt{Hor} (green) and \texttt{Ver} (blue) prediction modes for the 4$\times$4 block (red). Shaded regions are neighbouring feature values.}
\label{fig:prediction}
\vspace{-.4cm}
\end{figure}

Every 4$\times$4 block of feature values is predicted using one of five modes: palette (\texttt{Pal}), horizontal (\texttt{Hor}), vertical (\texttt{Ver}), and two filter modes (\texttt{Fil}). In the \texttt{Pal} mode, all values in the block are predicted using  $p_{j^{*}}$. %The $p_i$ that leads to fewest bits for the residual is selected and it is signalled to the decoder using the corresponding codeword in Table~\ref{tbl:mfp_table}. 
In \texttt{Hor}/\texttt{Ver} modes, the immediate left/top value is used as a predictor, as indicated in Fig.~\ref{fig:prediction}. If the block is at the left (top) boundary, $p_{j^{*}}$ is used as the left (top) value. %, and it is signalled to the decoder according to Table~\ref{tbl:mfp_table}. %For the boundary condition, the specified $P_{m}^{n}$ is used as a reference. $Pal$ mode predicts the all pels in a block with $P_{m}^{n}$. 
%Furthermore, it is employed to predict the all pels in the block in the $Pal$ mode. 
The two \texttt{Fil} modes are based on 3-tap filters with coefficients $[3, 7, 22]/32$ or $[14, 0, 18]/32$~\cite{alvar2016lossless}, and use the top-left, top, and left feature values to predict the current value. Again, at the boundaries, the unavailable values are replaced by $p_{j^{*}}$. %pels as references and multiply filter coefficients either $[3, 7, 22]/32$ or $[14, 0, 18]/32$~\cite{alvar2016lossless}. 

Prediction mode decision is based on the number of bits required for coding the residual, with the best mode being the one that requires least bits. %Therefore, the best prediction mode has the minimum bits cost. Note that distortion cost is not considered in lossless coding. 
In order to minimize the bits needed to specify the prediction mode, we exploit the \emph{most probable mode} (\texttt{mpm}) method~\cite{avc}, where \texttt{mpm} is derived from the previously-coded left, top-left and top blocks' prediction modes. The most frequently used mode among them is considered the \texttt{mpm}. If the current block's mode is the same as \texttt{mpm}, bit 1 is coded by CABAC~\cite{hevc_sze} to indicate it. Otherwise, bit 0 is coded, followed by two bits to indicate the mode\footnote{There are five prediction modes in total, so if the mode is not \texttt{mpm}, it must be one of the other four, which can be indicated by two bits.}.

Prediction residuals for each $4\times4$ block are coded by CABAC. The first bit is the SKIP indicator. If the residual is all-zero, the SKIP indicator is set to 1 and the encoder moves to the next block. Otherwise, the SKIP indicator is set to 0 and residuals are coded using one of  three scan orders: horizontal, vertical, and zig-zag. %  Each 4$\times$4 residual block has one bin to indicate all zero residual block. If the bin is equal to 1, then the block is skipped but the bin is only coded. Otherwise, coded residual data follow the bin(=0). %There are three types of scan order, horizontal, vertical and the zig-zag scan order. 
For the \texttt{Ver} (\texttt{Hor}) prediction mode, vertical (horizontal)  scan order is used. %\textcolor{red}{Is this so (\texttt{Ver} prediction implies horizontal scan?) Or is it the other way around?} 
Other modes use the zig-zag scan order. Locations of non-zero residuals are first indicated by binarizing the scanned block, with 1's placed at the locations of non-zero residuals and 0's placed elsewhere. %All residual pel locations are indicated by binary decision if non-zero residual exists at the location, the bin is equal to 1, otherwise 0. 
This binary vector is coded using CABAC. Finally, the non-zero residual values are coded in a manner similar to HEVC~\cite{hevc_sze}: values larger than 1 or 2 are flagged, the flags are CABAC-coded, and the non-flagged values are binarized using exponential Golomb-Rice coding, then coded by CABAC. %For the locations where have non-zero residual, we examine whether the non-zero value is larger than one or two. With regard to the decision, the flags are coded and the remained values that are not indicated by the flags are binarized using exponential Golomb-Rice coding, then coded by CABAC~\cite{hevc_sze}  

\section{Input reconstruction}
\label{sec:reconstruction}
Although the primary goal of collaborative intelligence is efficient inference, in some cases it may be desirable to also have the input image available in the cloud. For example, if the model detects an object of interest based on the features that were transmitted to the cloud, it might be useful to have the whole input image, which can then be stored or further processed in the cloud. The straightforward way is to simply send the whole input image from the mobile to the cloud, but this is not necessary, since a good approximation to the input image can be reconstructed from the transmitted features.  

To demonstrate this, we construct a \emph{mirror} model, indicated in the bottom right of Fig.~\ref{fig:proposal}, based on the the network in the mobile. Specifically, given the network in the mobile, the mirror model consists of the same number of layers, but in reverse order: convolutional layers from the mobile network are mapped to the convolutional layers with the same kernel size in the mirror model, while max-pooling layers from mobile network are mapped to up-sampling layers.

%, while max-pooling layers from mobile network are mapped to up-sampling layers. 

%In order to generate input source using the transferred data, we employ the mirrored front-end layers intuitively and simply. Unlike the well-known studies such as the generative adversarial network~\cite{goodfellow2014generative} or auto-encoder~\cite{kingma2013auto}, the proposed process exploits transferred deep features along with an independent network which replicates solely mirrored structure of the front-end and trains the network from scratch. We replace max-pooling layers with up-sampling layer, otherwise we use the convolutional layer. As sharing the transferred feature data, additional transmission cost is not required but occasionally computing power is necessary at the cloud server. 

The goal of the mirror model is to reconstruct the input image from the deep features transmitted to the cloud. We train the mirror model using a loss function that combines  structural similarity (SSIM)~\cite{ssim} and mean square error (MSE) between the input and the reconstructed image, as 
\begin{equation}
L = \lambda_{1} \cdot (1-\textup{SSIM}) + \lambda_{2} \cdot \textup{MSE}
\label{eq:loss}
\end{equation}
\noindent We used $\lambda_{1}=0.6$ and $\lambda_{2}=1$. The mirror model is trained from scratch using the Adam optimizer~\cite{adam} with the initial learning rate of $10^{-4}$.  A total  of 16,551 images from~\cite{pascal-voc-2007} and~\cite{pascal-voc-2012} are employed for training the model, with 20\% randomly selected as validation data and the remaining 80\% used for training. The test set consists of another 4,952  images from~\cite{pascal-voc-2007}. The maximum number of epochs is set to 50 and batch size to 32. The training stops when the validation loss starts increasing. %\textcolor{red}{We set the batch size of 32 and the stopped epochs vary depending on how deeper layers are trained. The maximum epochs are set to 50. The training session is stopped when the loss cost starts increasing for the validation set.} 
%The training set consists of 16,551 images from~\cite{pascal-voc-2007} and~\cite{pascal-voc-2012}. 
%The reason that the structure dissimilarity is involved in the loss function is to consider structural characteristics of the reconstructed image along with more straightforward metric, MSE. 

\begin{figure*}[t]
    \begin{minipage}[b]{1.0\linewidth}
    \centering
    \includegraphics[width=\textwidth]{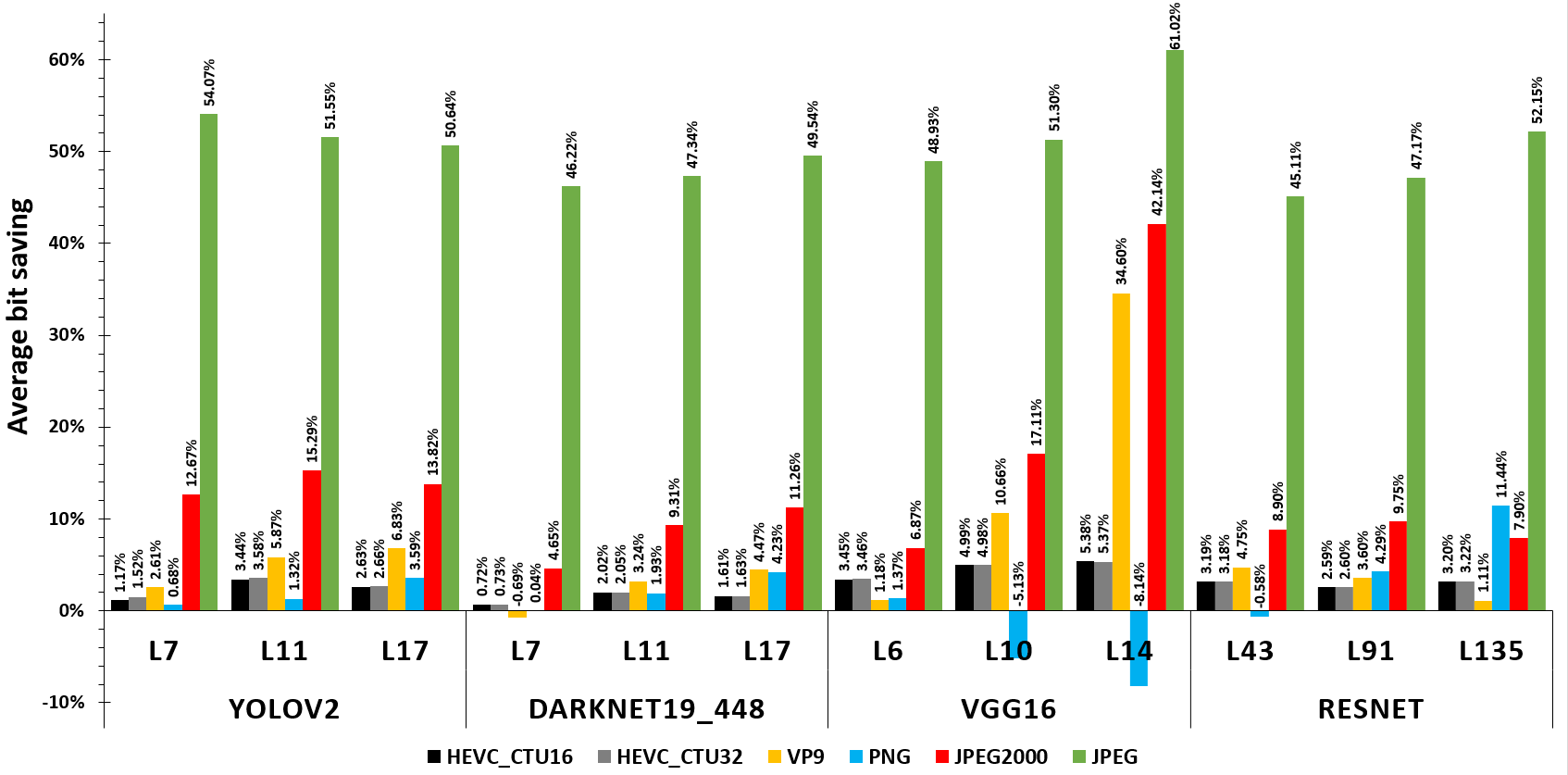}
    \end{minipage}
%\vspace{-0.6cm}
\caption{Comparison of average bits saving for the three different deep features from each network against conventional compression algorithms}
\label{fig:compression_performance}
\vspace{-.4cm}
\end{figure*}

\section{Experiments}
\label{sec:experiments}
The proposed deep feature compression was tested on four deep models: YOLOv2~\cite{YOLO2}, Darknet19\_448~\cite{darknet}, VGG16~\cite{vgg16} and ResNet~\cite{resnet}. YOLOv2 is a state-of-the-art object detector, while other models are used for image classification. %First, we look into the each networks' structure, then properly select the intermediate layers which produce the feature data to compress. In general, the data volume of the selected layer's output is less than or at least similar to the input data volume by the one of strategies in the collaborative approach~\cite{kang2017neurosurgeon, jointdnn, dfc_for_collab_object_detection}. As shown in the
Table~\ref{tbl:layer_structure} shows the size of the feature tensor at the output of three layers from each of the models, along with the dimensions of the feature matrix after tiling. %, we choose three layers from each network. The column of the layer output presents number of channels, channel width and height in order. In order to compress we tile the each channel within a frame and the frame size is shown in the last column. 
For testing compression of YOLOv2 features we used 4,952 images from VOC2007~\cite{pascal-voc-2007}, and for other models we used 50,000 images from ImageNet~\cite{ILSVRC15}.

Deep features produced by the various models were compressed using the proposed method, as well as the lossless versions of  %Now, the acquired deep feature data from the each network with proper data input are compressed by the proposed compressor and also by five benchmark compressors, 
HEVC~\cite{hevc}, VP9, PNG, JPEG2000 and JPEG. %Especially, HEVC has variety of the configuration, but 
For HEVC, we followed common test conditions~\cite{hevc_ctc} associated with lossless coding, while changing the largest coding unit size, also known as CTU, to 32$\times$32 and 16$\times$16. %For the rest of the codecs we employ  FFMPEG~\footnote{https://www.ffmpeg.org/} and it also supports lossless compression mode. 

As usual, evaluation of lossless coding is based on the number of bits used. %Due to the fact that we compress the data losslessly, evaluation metric computes average bits saving for the proposed algorithm compared to the benchmark.  
Fig.~\ref{fig:compression_performance} shows the average bit difference between the proposed method and each of the five competing methods, on the image datasets described above. Positive values mean that the competing method uses more bits than the proposed one. As seen in the figure, HEVC (with both CTU sizes) needs 0.7-3.0\% more bits than the proposed method in most cases, and up to 5\% more for the tenth and fourteenth layer of VGG16. %Note that compression performance gap between two different size of CTU is negligible.
VP9 also uses more bits than the proposed method (up to 34\% more in the fourtenth layer of VGG16), except for the seventh layer of the Darkent19\_448 where it uses 0.96\% fewer bits. PNG turns out to be a very good codec for deep feature data. While it uses more bits than the proposed method in most cases, it needs up to 9\% fewer bits in the fourteenth layer of VGG16. %is a straightforward lossless codec using dictionary coder with simple structure. In some sense, a way of doing prediction by PNG could have similarity with the proposed method. The blue bars in the fig.~\ref{fig:compression_performance} indicates compression performance against PNG coder. Most cases the proposed saves average bits up to 11.4\%, but there are two significant increase in the tenth and fourteenth layer in VGG16~\cite{vgg16}. 
Both JPEG2000 and JPEG require considerably more bits than other codecs. Compared to the proposed method, JPEG2000 needs up to 45\% more bits and JPEG needs up to 61\% more bits. % (red bars) shows that it need noticeably more bits to compress deep feature data compared to the proposed method for all cases. Compared to the very commonly used JPEG(gree bars), remarkably the proposed method saves average bits at least 45\% up to 61\%. 
Also, Table~\ref{tbl:mode_sel} shows the average mode selection percentage for each model in the study. As observed in Table~\ref{tbl:mode_sel}, \texttt{Pal} is the most frequently used prediction mode in each of the four models, accounting for 35-58\% of predictions.

\begin{table}[b]
\centering
\vspace{-0.3cm}
\caption{Dimensions of feature tensors and feature matrices at three different layers of each of the models in the study.}
\label{tbl:layer_structure}
%\smallskip\noindent
%\resizebox{\linewidth}{!}{%
\begin{tabular}{@{}cccc@{}}
\toprule
Model                         & Layer & Feature tensor & Feature matrix \\ \midrule
\multirow{3}{*}{YOLOv2}       & L7        & 128$\times$52$\times$52    & 832$\times$416    \\
                                & L11       & 256$\times$26$\times$26    & 416$\times$416    \\
                                & L17       & 512$\times$13$\times$13    & 416$\times$208    \\ \midrule
\multirow{3}{*}{Darknet19\_448} & L7        & 128$\times$56$\times$56    & 896$\times$448    \\
                                & L11       & 256$\times$28$\times$28    & 448$\times$448    \\
                                & L17       & 512$\times$14$\times$14    & 448$\times$224    \\ \midrule
\multirow{3}{*}{VGG16}          & L6        & 128$\times$56$\times$56    & 896$\times$448    \\
                                & L10       & 256$\times$28$\times$28    & 448$\times$448    \\
                                & L14       & 512$\times$14$\times$14    & 448$\times$224    \\ \midrule
\multirow{3}{*}{ResNet}         & L43       & 128$\times$32$\times$32    & 512$\times$256    \\
                                & L91       & 256$\times$16$\times$16    & 256$\times$256    \\
                                & L135      & 256$\times$16$\times$16    & 256$\times$256    \\ \bottomrule
\end{tabular}
\vspace{-0.5cm}
\end{table}

\begin{table}[t]
\centering
\caption{Average mode selection percentage for each model}
\label{tbl:mode_sel}
\begin{tabular}{@{}cccccc@{}}
\toprule
Model    & \texttt{Hor}   & \texttt{Ver}   & \texttt{Pal}   & \texttt{Fil1}  & \texttt{Fil2} \\ \midrule
YOLOv2  & 31.67 & 18.10 & 35.02 & 9.95  & 5.26 \\
Darknet19\_448 & 13.56 & 13.91 & 58.06 & 9.21  & 5.26 \\
VGG16   & 25.14 & 19.92 & 42.93 & 7.83  & 4.18 \\
ResNet  & 17.25 & 12.80 & 51.42 & 14.13 & 4.40 \\ \bottomrule
\end{tabular}
\vspace{-.5cm}
\end{table}

\begin{figure*}[t]
    \begin{minipage}[b]{1.0\linewidth}
    \centering
    \includegraphics[width=\textwidth]{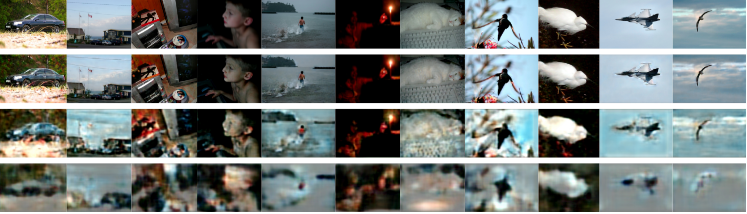}
    \end{minipage}
%\vspace{-0.6cm}
\caption{Top row: original input images. Other rows: reconstructed images from the seventh, eleventh, and seventeenth layer of YOLOv2.}
%\vspace{-0.3cm}
\label{fig:inference_input}
\vspace{-.3cm}
\end{figure*}

Finally, we demonstrate input reconstruction from the features generated at the seventh, eleventh and seventeenth layer of YOLOv2. Hence, three mirror models are trained for reconstruction, one for each set of features. 
Table~\ref{tbl:input_inference_results} shows the average Peak Signal to Noise Ratio (PSNR, in dB) and SSIM, along with standard deviations, over the test set. % index with variances against original inputs. In general, chrominance inference shows better quality than luminance component. A common observation of the input inference process is 
As seen in the table, the deeper the layer from which features are extracted, the more difficult is the input reconstruction, since more information gets lost in max-pooling layers. % feature data, the more difficult it is to deduce the original source since the deeper feature data typically passes through the more number of max-pooling layer which is a non-linear function. 
Visual results, shown in Fig.~\ref{fig:inference_input}, look somewhat better than what is suggested by quantitative results in Table~\ref{tbl:input_inference_results}. The first row shows the original input images, while the remaining rows show reconstructed images from the seventh, eleventh and seventeenth layer, in that order. Reconstructions from the seventh layer look reasonably good compared to the original images. However, reconstructions from deeper layers start to lose important details.

\begin{table}[b]
\centering
\vspace{-0.3cm}
\caption{Comparison of PSNR \& SSIM index for the reconstructed input images compared to original inputs.}
\label{tbl:input_inference_results}
%\smallskip\noindent
%\resizebox{\linewidth}{!}{%
\begin{tabular}{@{}llllll@{}}
\toprule
                     &   & \multicolumn{2}{c}{PSNR}                          & \multicolumn{2}{c}{SSIM}                          \\ \midrule
                     &   & \multicolumn{1}{c}{Avg.} & \multicolumn{1}{c}{Std.} & \multicolumn{1}{c}{Avg.} & \multicolumn{1}{c}{Std.} \\ \midrule
\multirow{3}{*}{L7}  & Y & 23.8633     & 2.8674     & 0.7042      & 0.1107     \\
                     & U & 35.8363     & 3.8942    & 0.9411      & 0.0344     \\
                     & V & 33.0488     & 4.1624    & 0.9160      & 0.0452     \\ \midrule
\multirow{3}{*}{L11} & Y & 17.5348     & 1.7231     & 0.4743      & 0.1319     \\
                     & U & 30.7260     & 2.6935     & 0.9106      & 0.0466     \\
                     & V & 27.6556     & 3.1462     & 0.8645      & 0.0613     \\ \midrule
\multirow{3}{*}{L17} & Y & 14.2898     & 1.7578     & 0.4044      & 0.1367     \\
                     & U & 28.6288     & 3.5730    & 0.9072      & 0.0505     \\
                     & V & 25.8074     & 3.9733    & 0.8560      & 0.0648     \\ \bottomrule
\end{tabular}
\vspace{-0.5cm}
\end{table}

%\subsubsection{Subsubsection Heading Here}
%Subsubsection text here.

\section{Conclusion}
\label{sec:conclusion}
In this study, we examined the characteristics of deep feature data and proposed a simple and effective method for near-lossless deep feature compression. The proposed method outperforms state-of-the-art image codec in this regard. We also demonstrated input image reconstruction from deep feature data by constructing and training a mirror model. Future work will involve the development of lossy compression schemes for deep feature data.  %, the output of the intermediate layer of the neural network, in order to improve transmission efficiency by improving compression ratio in collaborative approach. Based on the examination, we proposed a deep feature compression algorithm which is simple, but outperforms the state-of-the-art compression algorithm for different networks deep features. Furthermore, we suggested the input data inference process in the cloud server. Without the need for data, the process predicts input data by passing the the transmitted deep features through the mirrored front-layers. The predicted inputs shows proper quality compared to the original input sources.

\bibliographystyle{IEEEbib}
\bibliography{ref}

\begin{thebibliography}{10}

\bibitem{poniszewska2018endowing}
A.~Poniszewska-Maranda, D.~Kaczmarek, N.~Kryvinska, and F.~Xhafa,
\newblock ``Endowing iot devices with intelligent services,''
\newblock in {\em Proc. Int. Conf. Emerging Internetworking, Data \& Web
  Technol.}, 2018, pp. 359--370.

\bibitem{kang2017neurosurgeon}
Y.~Kang, J.~Hauswald, C.~Gao, A.~Rovinski, T.~Mudge, J.~Mars, and L.~Tang,
\newblock ``Neurosurgeon: Collaborative intelligence between the cloud and
  mobile edge,''
\newblock in {\em Proc. 22nd ACM Int. Conf. Arch. Support Programming Languages
  and Operating Syst.}, 2017, pp. 615--629.

\bibitem{jointdnn}
A.~E. Eshratifar, M.~S. Abrishami, and M.~Pedram,
\newblock ``{JointDNN}: an efficient training and inference engine for
  intelligent mobile cloud computing services,''
\newblock {\em arXiv preprint arXiv:1801.08618}, 2018.

\bibitem{dfc_for_collab_object_detection}
H.~Choi and I.~V. Baji\'{c},
\newblock ``Deep feature compression for collaborative object detection,''
\newblock in {\em Proc. IEEE ICIP'18}, 2018,
\newblock to appear.

\bibitem{YOLO2}
J.~Redmon and A.~Farhadi,
\newblock ``{YOLO9000:} better, faster, stronger,''
\newblock in {\em Proc. IEEE CVPR'17}, Jul. 2017, pp. 6517--6525.

\bibitem{hevc_complexity_analysis}
F.~Bossen, B.~Bross, K.~Suhring, and D.~Flynn,
\newblock ``{HEVC} complexity and implementation analysis,''
\newblock {\em IEEE Trans. Circuits Syst. Video Technol.}, vol. 22, pp.
  1685--1696, Dec. 2012.

\bibitem{deepsic}
S.~Luo, Y.~Yang, and M.~Song,
\newblock ``{DeepSIC}: Deep semantic image compression,''
\newblock {\em arXiv preprint arXiv:1801.09468}, 2018.

\bibitem{CABAC_H264}
D.~Marpe, H.~Schwarz, and T.~Wiegand,
\newblock ``Context-based adaptive binary arithmetic coding in the {H.264/AVC}
  video compression standard,''
\newblock {\em IEEE Trans. Circuits Syst. Video Technol.}, vol. 13, pp.
  620--636, July 2003.

\bibitem{vgg16}
A.~Zisserman K.~Simonyan,
\newblock ``Very deep convolutional networks for large-scale image
  recognition,''
\newblock {\em arXiv preprint arXiv:1409.1556}, 2014.

\bibitem{Goodfellow-et-al-2016}
G.~Ian, B.~Yoshua, and C.~Aaron,
\newblock {\em Deep Learning},
\newblock MIT Press, 2016.

\bibitem{pascal-voc-2007}
M.~Everingham, L.~Van~Gool, C.~K.~I. Williams, J.~Winn, and A.~Zisserman,
\newblock ``The {PASCAL} {V}isual {O}bject {C}lasses {C}hallenge 2007
  {(VOC2007)} {R}esults,'' http://host.robots.ox.ac.uk/pascal/VOC/voc2007/.

\bibitem{sze2014entropy}
V.~Sze and D.~Marpe,
\newblock ``Entropy coding in {HEVC},''
\newblock in {\em High Efficiency Video Coding (HEVC)}, pp. 209--274. Springer,
  2014.

\bibitem{alvar2016lossless}
S.~R. Alvar and F.~Kamisli,
\newblock ``On lossless intra coding in hevc with 3-tap filters,''
\newblock {\em Signal Processing: Image Comm.}, vol. 47, pp. 252--262, 2016.

\bibitem{avc}
T.~Wiegand, G.~J. Sullivan, G.~Bjontegaard, and A.~Luthra,
\newblock ``Overview of the {H.264/AVC} video coding standard,''
\newblock {\em IEEE Trans. Circuits Syst. Video Technol.}, vol. 13, pp.
  560--576, Jul. 2003.

\bibitem{hevc_sze}
V.~Sze and M.~Budagavi,
\newblock {\em High Efficiency Video Coding (Algorithms and Architectures)},
\newblock Springer, 2014.

\bibitem{ssim}
Z.~Wang, L.~Lu, and A.~C. Bovik,
\newblock ``Video quality assessment based on structural distortion
  measurement,''
\newblock {\em Signal processing: Image Comm.}, vol. 19, no. 2, pp. 121--132,
  2004.

\bibitem{adam}
D.~P. Kingma and J.~Ba,
\newblock ``Adam: A method for stochastic optimization,''
\newblock in {\em Proc. ICLR'15}, 2015.

\bibitem{pascal-voc-2012}
M.~Everingham, L.~Van~Gool, C.~K.~I. Williams, J.~Winn, and A.~Zisserman,
\newblock ``The {PASCAL} {V}isual {O}bject {C}lasses {C}hallenge 2012
  {(VOC2012)} {R}esults,'' http://host.robots.ox.ac.uk/pascal/VOC/voc2012/.

\bibitem{darknet}
J.~Redmon,
\newblock ``{Darknet: Open source neural networks in C.},''
  http://pjreddie.com/darknet/, 2013-2017,
\newblock Accessed: 2017-10-19.

\bibitem{resnet}
K.~He, X.~Zhang, S.~Ren, and J.~Sun,
\newblock ``Deep residual learning for image recognition,''
\newblock in {\em Proc. IEEE CVPR'16}, 2016, pp. 770--778.

\bibitem{ILSVRC15}
O.~Russakovsky, J.~Deng, H.~Su, J.~Krause, S.~Satheesh, S.~Ma, Z.~Huang,
  A.~Karpathy, A.~Khosla, M.~Bernstein, A.~C. Berg, and Li~F,
\newblock ``{ImageNet Large Scale Visual Recognition Challenge},''
\newblock {\em Int. Journal of Computer Vision}, vol. 115, no. 3, pp. 211--252,
  2015.

\bibitem{hevc}
G.~J. Sullivan, J.-R. Ohm, W.-J. Han, and T.~Wiegand,
\newblock ``Overview of the high efficiency video coding {(HEVC)} standard,''
\newblock {\em IEEE Trans. Circuits Syst. Video Technol.}, vol. 22, no. 12, pp.
  1649--1668, 2012.

\bibitem{hevc_ctc}
F.~Bossen,
\newblock ``Common {HM} test conditions and software reference
  configurations,''
\newblock in {\em ISO/IEC JTC1/SC29 WG11, {JCTVC-L1100}}, Jan. 2013.

\end{thebibliography}

\end{document}